# Core Electron Spectroscopic Studies for 11-Fe-based Superconductors

Soumyadeep Ghosh[a,b] *, Haranath Ghosh[a,b]1

[a] *Human Resources Development Section, Raja Ramanna Centre for Advanced Technology, Indore 452013, India*
[b] *Homi Bhabha National Institute, 2nd floor, Training School Complex, Anushakti Nagar, Mumbai 400094, India*

**Abstract**

Electron energy loss near edge spectra (ELNES) has been calculated for two different Fe based superconducting material FeSe and FeTe using density functional theory (DFT). Fe K-edge absorption spectra consist of several features, origins of which are thoroughly described in light of partial density of states of constituent atoms. Here we have included "core-hole effect" and found drastic change in absorption spectra. Our results including core hole effect matches very well with experimental X-ray Absorption Near Edge Structure (XANES) results available in literature.

**1. Introduction**

High temperature superconductivity in iron based superconducting materials had been one of the frontier research area since last 10 years. There exist a large number of families of Fe-based superconducting materials, FeSe, FeTe and their variants belong to the simplest structured 11 family, in which Fe atoms forms a square lattice and Se/Te are above and below the Fe-plane. Even though 11 family members have the simplest structures among all other families [1], it holds the record highest Tc ~ 100 K [2] when grown on single layered $SrTiO_3$. Both Electron energy loss near edge structure (ELNES) and x-ray absorption near edge structure (XANES) spectroscopy measure the electric dipole transitions from a selected core-orbital to unoccupied states. Due to the high energy resolution of EELS spectrometer, it can provide information on bonding, local composition, structure, site occupancy and electronic structure of materials. The probing electron or X-ray (depending on ELNES or XANES studies) interacts with the system (like atom/molecule/solid) which thus gets perturbed by the probe itself. The system being studied will have an inner shell electron removed, leaving a core hole, in general known as "core-hole" effect. The core hole effect, essentially have a net positive charge of the target atom, provides a pull to the unoccupied density of states towards the conduction band edge, causes redistribution of the unoccupied electronic density of state. Such studies in Fe-based superconducting materials provided a special role [3,4]. In this work we provide Fe-K edge absorption features for the two Fe-based superconducting materials FeSe and FeTe within first principles density functional theory (DFT) framework. Our results show excellent agreement with experimental observations.

**2. Computational Methodology**

DFT based simulations of ELNES and calculation of partial density of states (PDOS) are performed using CASTEP [5], which is based on the plane-wave pseudo-potential method [6]. Exchange correlation is taken as generalized gradient approximation (GGA) using using Perdew-Burke-Enzerhof (PBE) functional [7]. Here we use

---

\* Corresponding author. Tel.: 0731244-2580
  E-mail address: soumyadeep@rrcat.gov.in

BFGS geometry optimization scheme [8] for fully relaxed structure. We used experimentally determined crystal structure for FeSe having tetragonal Bravais lattices with space group P4/nmm (No.-129) [9].

## 3. Results and Discussions

In Fig. (1) theoretically calculated Fe K-edge absorption spectra for two iron based superconductors FeSe (Fig.-1:i) and FeTe (Fig.-1:ii) with and without including core hole effect is presented. In case of FeSe there are five distinct features in ELNES spectra (A, B, C, D and E). The feature A is known as 'pre-edge' feature which signifies amount of distortion of $FeSe_4$ / $FeTe_4$ tetrahedrons. Feature A is due to transition from *1s* to *3d*, which is known to be very weak due to quadrupolar nature of transition. However due to Fe(Se/Te)$_4$ distortion there are mixing between *2p-3d* orbitals of the legends and transition metals. Due to such mixing of different angular momentum electronic states, *1s* to *3d* optical transition gains moment and shows in the core-level excited spectra. This indicates that FeSe has larger tetrahedral distortion compared to FeTe. Such mixing between *2p-3d* electronic states have been elaborately shown in the third row figures of Figs. (2,3:iii). The features B, C, D are due to *1s* to *2p* transitions where the *2p* state can come either from Se/Te or Fe as distinctly presented in Figs. (2, 3). However, modifications in the core-level excited spectra are worth noticing in presence of core-hole effect. Feature B becomes most intense among all others differing so much from no core hole spectra in which feature C is most intense. Feature C is suppressed with inclusion of core hole and shifted toward higher energy region. This is because the feature D is enhanced largely and spectral weight is transferred. Feature D and E both are also significantly enhanced, become more intense and wider by the inclusion of core hole without any horizontal shift in loss energy. In case of FeTe we get four distinct features (A, B, C and D). Like FeSe feature A, B becomes pronounced with inclusion of core hole. Feature C shifts towards low energy region as well as suppressed and feature D becomes pronounced due to core hole effect, also width of both feature increases.

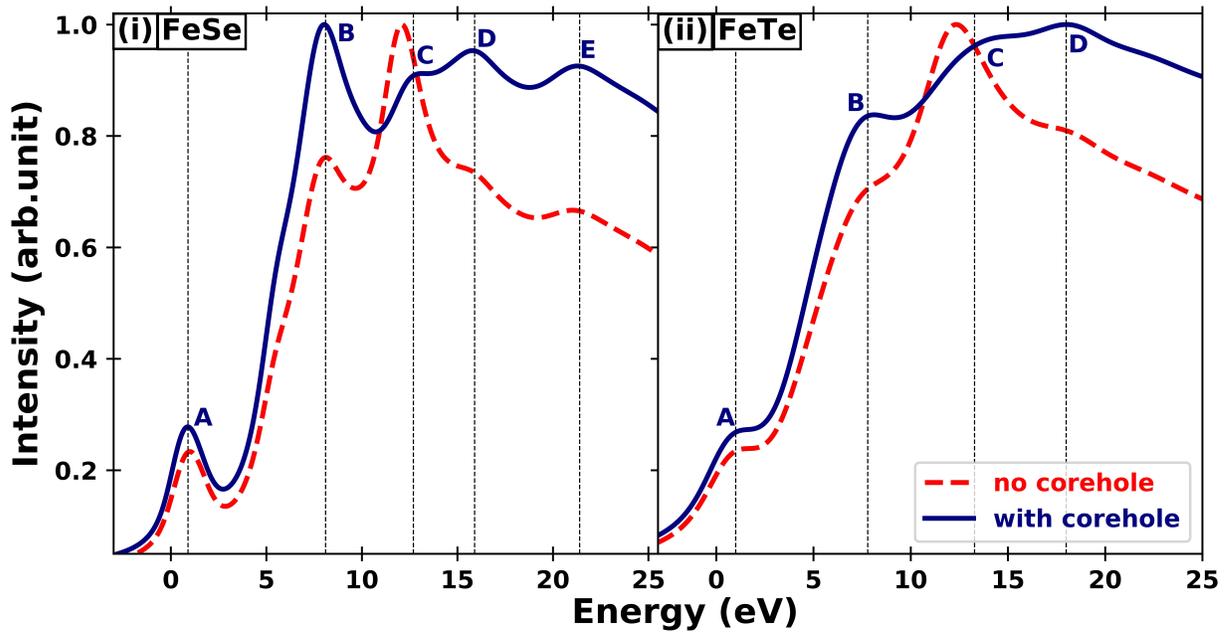

Fig. 1. Normalized Fe K-edge absorption spectra with (solid lines) and without (dashed lines) core hole effects for FeSe (left), FeTe (right) respectively. Special features are indicated by vertical lines.

## 4. Conclusions

From the above results we can conclude that core hole effect plays significant role in determination of Fe K-edge absorption spectra for FeSe and FeTe. For FeSe pre-edge feature is prominent and pronounced in presence of core hole effect. In case of FeTe pre-edge feature is less prominent than FeSe even after considering core hole effect. From electronic structure point of view transition from *1s* state to any empty *p* state is most intense since dipolar transitions are dominant. All the main peaks and separation between different peaks including core hole effect are in close agreement with experimental observations [3,10].

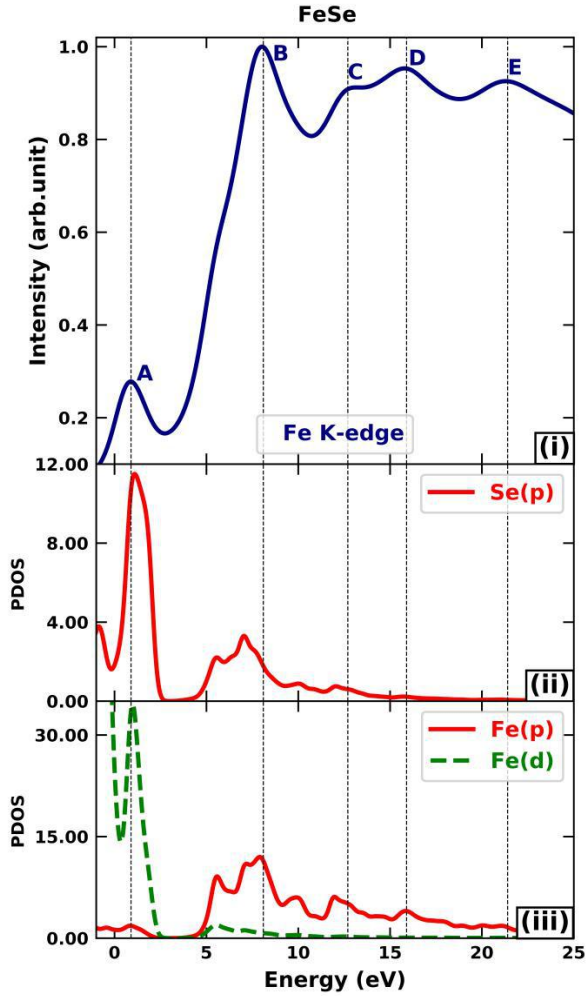

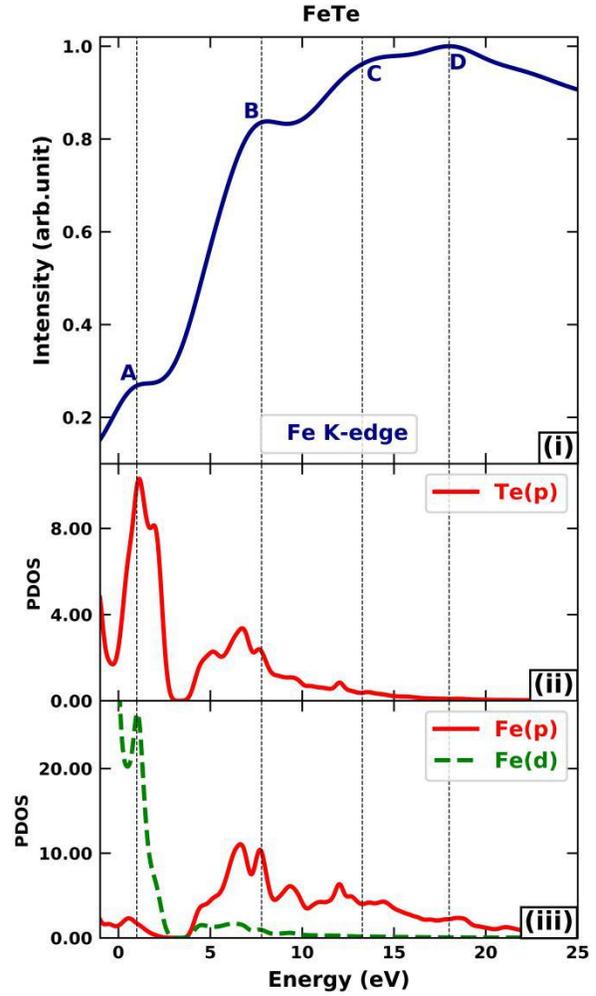

Fig. 2. Theoretical ELNES spectra of FeSe: Fe K-edge absorption spectra (i) and site projected PDOS (ii-iii)

Fig. 3. Theoretical ELNES spectra of FeTe: Fe K-edge absorption spectra (i) and site projected PDOS (ii-iii)

**Acknowledgements**

We thank Dr. P. A. Naik for his encouragement and A. Ghosh for his involvement during this work. HG and SG acknowledges RRCAT computer center for support. SG acknowledges the HBNI, RRCAT for financial assistance.